\documentclass[10pt,conference,twocolumn]{IEEEtran}

\usepackage{threeparttable}

\usepackage{amsmath}
\usepackage[noadjust]{cite}
\usepackage{enumerate}
\usepackage{amssymb}
\usepackage{subcaption}
\usepackage{graphicx}
\usepackage{float}
\usepackage{epstopdf}
\usepackage{pbox}
\usepackage{color}
\usepackage[section,subsection,subsubsection]{extraplaceins}
\usepackage{multirow}
\usepackage{booktabs}
\usepackage{citesort}

\newcommand{\RNum}[1]{\uppercase\expandafter{\romannumeral #1\relax}}
\newcommand{\chanaka}[1]{\color{brown} }
\usepackage{arydshln}
\usepackage{threeparttable}
\usepackage[normalem]{ulem}

\usepackage[cmintegrals]{newtxmath}
\usepackage{algorithmic}
\usepackage{array}

\begin{document}

\title{{On the Benefits of Multi-hop Communication for Indoor 60~GHz Wireless Networks}}


{\author{\IEEEauthorblockN{Chanaka Samarathunga\IEEEauthorrefmark{1}, Mohamed Abouelseoud\IEEEauthorrefmark{2}, Kazuyuki Sakoda\IEEEauthorrefmark{3}, Morteza Hashemi\IEEEauthorrefmark{1}}\IEEEauthorblockA{\IEEEauthorrefmark{1}Department of Electrical Engineering and Computer Science, University of Kansas \\ \IEEEauthorrefmark{2}Sony R\&D Center US, San Jose lab, \IEEEauthorrefmark{3}Sony R\&D Center Japan, Tokyo lab}}}

\maketitle

\begin{abstract}
 The spectrum-rich millimeter wave (mmWave)  frequencies 
have the potential to alleviate the spectrum crunch that the wireless and cellular operators are already experiencing. However, compared with traditional wireless communication in the sub-6 GHz bands, due to small
wavelengths most objects such as human body, cause significant additional path losses (up to 20 dB), which can entirely break the mmWave link. Also, mmwave links suffer from limited range of communication. In this paper, we resort to network layer solutions to demonstrate the benefits of multi-hop routing in mitigating the blockage issue and extending communication range in mmWave band.  To this end, we develop a hop-by-hop multi-path routing protocol that finds one primary and one backup next-hop per destination in order to guarantee reliable and robust communication under extreme stress conditions. System-level simulations based on the IEEE 802.11ad specifications demonstrate that the proposed routing protocol provides a reliable end-to-end throughput performance, while satisfying the latency requirements.
\end{abstract}
\section{Introduction}
   
Due to the emerging applications and the need for higher capacity, current sub-6 GHz wireless spectrum is not enough to cope with the recent demands for high data rate applications ranging from augmented/virtual reality to wireless HD video streaming. Millimeter wave (mmWave) bands -- between 30 GHz to 300 GHz -- have been contemplated as a solution to mitigate the existing spectrum scarcity in the sub-6 GHz. However,  enabling mmWave wireless systems in general requires properly dealing with the channel impairments and propagation characteristics of the high frequency bands. In particular, due to the high propagation losses, mmWave provides a limited range of communications. Also, high penetration, reflection and diffraction losses reduce the available diversity and limit non-line-of-sight (NLOS) communications. Due to the small wavelength of mmWave, most objects in the propagation environment lead to blocking and reflection as opposed to scattering and diffraction as in the sub-6 GHz bands.  

\begin{figure}[t]
	\centering
	\includegraphics[scale=.24,trim=.5cm 0.3cm 1cm 1cm, clip]{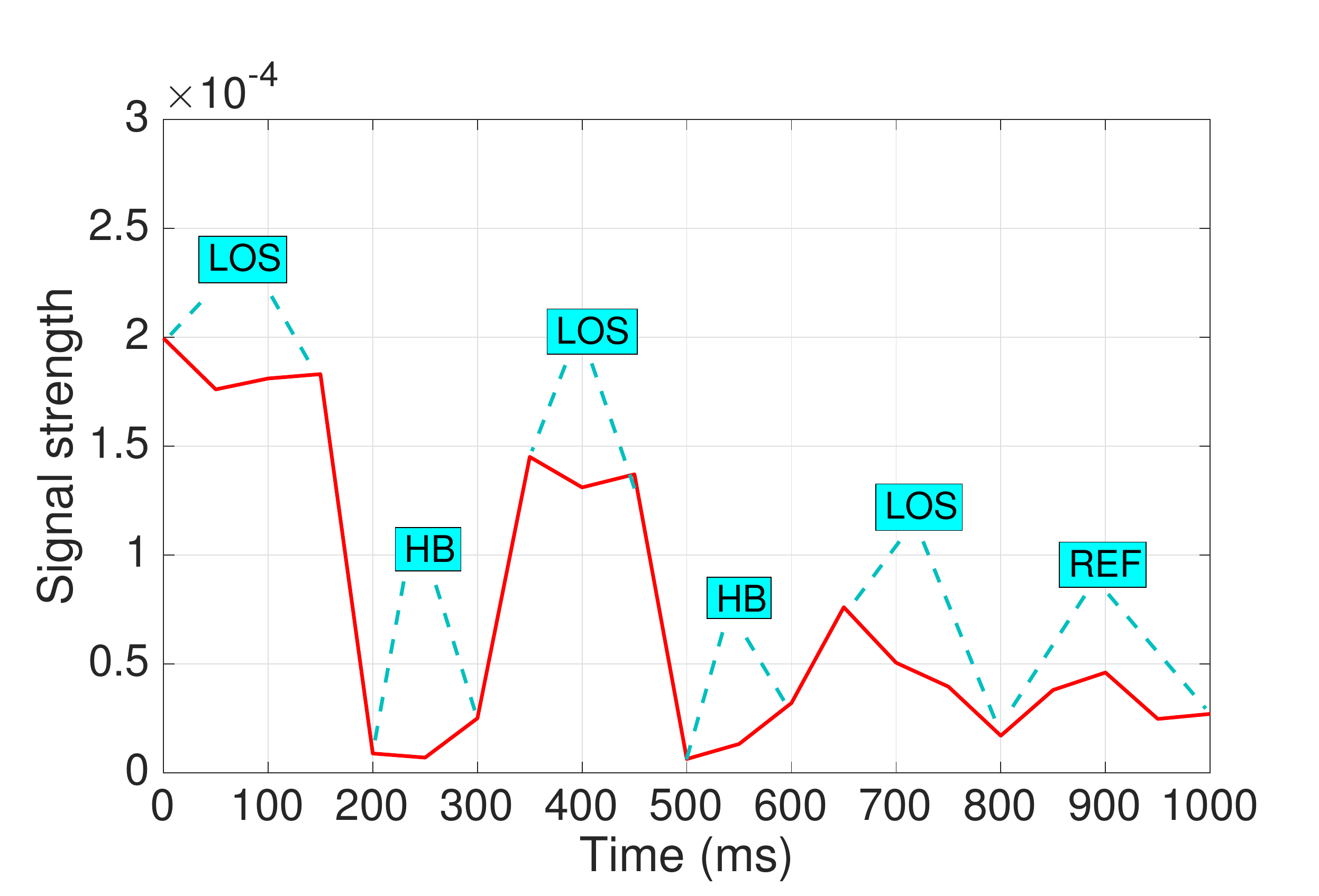}
	\caption{\small{Received mmWave signal strength under line of sight
			(LOS), human blocker (HB), and reflection (REF) \cite{hashemi2018out}.}}
	\label{blockageexperiment}
\end{figure}

To experimentally investigate the effect of human blockage on mmWave links, we conducted a set of measurements in the 30 GHz band with a stationary transmitter and a mobile receiver that moves away from the transmitter with the speed of 1 m/s~\cite{hashemi2018out}. From the results shown in Fig. \ref{blockageexperiment}, we observe that during human blockage intervals (denoted by ``HB"), the received signal strength falls to almost zero. This is consistent with the measurement results in \cite{slezak2018empirical} that suggest human body can increase the path loss by more than 20 dB.



Currently, there is arguably an adequate understanding of physical layer issues and signal characterization \cite{rappaport2013millimeter,rangan2014millimeter} as well as MAC layer designs for mmWave systems \cite{shokri2016design}. By contrast, the upper layers of the protocol stack are still largely unexplored and the existing protocols are not tailored for the mmWave bands.  Multi-hop networking can be a viable solution to mitigate the issue of limited range of communications as well as the sensitivity to blockage. Notwithstanding the benefits, multi-hop communication adds additional overhead by requiring exchange of metadata (e.g., route discovery messages), which is more challenging due to directional communication in mmWave bands. In addition, it is not clear that multi-hop networking satisfies the latency requirements due to  extra processing, queuing, and channel access latency at relay nodes. For instance, the IEEE 802.11ay use-case document specifies the latency and jitter requirements of less than 5ms for 8K UHD wireless transfer at smart homes as well as for augmented reality/virtual reality headsets and other high-end wearable devices \cite{802.11ay}.   

In this paper, we investigate the performance of multi-hop mmWave communication for indoor 60 GHz applications and demonstrate the benefits of multi-hop routing protocols for blockage-prone and range-limited mmWave links.  Due to the nature of use-cases for indoor mmWave systems (e.g., real-time high data rate applications), a blocked link should quickly be detected and replaced by an alternative link. As such, we propose a hop-by-hop multi-path routing protocol that is efficient and fast in switching to a \emph{reserved ready-to-use path} towards the destination. We implement the proposed protocol on top of the 802.11ad PHY and MAC specification, and investigate its performance under different scenarios. 


\section{Related Work}
\label{sec:related_work}
There are two main categories of related works: (i) On-demand routing protocols that include the basic Ad-hoc On-Demand Distance Vector (AODV), multi-path AODV, and directional multi-hop protocols, and (ii) methods to combat blockage in mmWave bands. 

\subsection{On-Demand Routing Protocols}
\textbf{Basic AODV:} In a typical multi-hop network, a route from the originating STA to the destination STA is determined by selecting intermediate STAs for the end-to-end path. Often, the intermediate STAs are chosen so that the links for use offer best link quality. 
AODV is an on-demand routing protocol that represents the general essence of multi-hop routing over wireless media. With AODV, STAs generate routes based on flooding route request (RREQ) and route reply (RREP) messages. The originating STA broadcasts RREQ frames and intermediate STAs receive the RREQ frames, measure quality of the link between itself and originating STA, and rebroadcast the RREQ embedding the link quality information. This process continues until the destination STA receives RREQ frames.
As a result, 
the destination STA finds a route with a better link quality to the originating STA. Next, the destination STA sends back RREP frames to confirm the best route to intermediate and originating STAs. 
The originating STA receives RREP and establishes a multi-hop route toward the destination STA. 

\textbf{Multi-path AODV:} An AODV backup routing scheme to create a mesh structure and provide multiple alternate routes has been proposed in \cite{Lee2000AODV-BR}. However, this scheme is not designed for heavy traffic networks. In \cite{LAI2007453}, AODV-ABR (Adaptive Backup Route) scheme and AODV-ABL scheme  which is the combination of AODV-ABR and local repair are proposed. The study in \cite{Khare2013} has proposed a low-latency AODV routing protocol for mobile ad-hoc network (MANET) by developing route failure prediction mechanism. When the node in primary route detects that received signal strength is less than a threshold value it switches to the alternate route. This would eliminate the route rediscovery process by the source upon a route failure. A cross-layered multi-path AODV (CM-AODV) protocol which selects
multiple routes on demand based on the SINR measured at the
physical layer has been proposed in \cite{park2008}. In this protocol the RREQ packets transmitted during route
discovery are broadcast packet. An improved multi-channel AODV routing protocol based on Dijkstra algorithm has been proposed in \cite{Liu2019}. The proposed protocol uses two channels to transmit data packets and control messages separately. Path quality has been evaluated using packet reception rate and Dijkstra algorithm is used to find the optimal path. An energy conserving routing protocol for MANETs, named EM-AODV (Energy Multi-path AODV) is proposed in \cite{khelifa2010}.
These protocols have not been designed and tested for  mmWave systems, where, for example, due to the need for directional transmissions, broadcasting is not feasible for flooding route discovery messages. 

\textbf{Directional multi-hop protocols:} There are several works proposing directional MAC protocols for multi-hop wireless networks \cite{singh,4133898,niu2014blockage,gossain2006drp}. The authors in \cite{singh} have proposed slotted ALOHA-based protocol for ad-hoc networks with devices using adaptive array smart antennas, and  \cite{4133898} has proposed polling based MAC protocol in order to discover and track its neighbors. While these protocols are designed for directional systems, they are not tailored for mmWave and do not address frequent route break scenarios that happen more frequently in mmWave bands compared to sub-6 GHz frequencies. mmWave use cases require extremely high data rates with low latency, which in turn calls for agile multi-hop routing protocols that can quickly react to broken links. 

\subsection{Methods to Combat mmWave Blockage}
In order to mitigate the blockage issues in mmWave systems, there are several physical and MAC layer proposals such as exploiting reflection paths from walls \cite{genc2010robust,yiu2009empirical}, using intelligent reflecting surfaces \cite{wu2018intelligent,qingqing2019towards}, using different antenna gains \cite{8608384}, and  integrating mmWave with lower frequencies \cite{yao2019integrating,hashemi2018out}.
There are also some works proposing deflection routing schemes for throughput improvement and mitigating blockage issues~\cite{4917660,5378881}. The source device finds a relay and creates a deflection route when there is a blockage in direct path \cite{5378881}. In this method, an alternating route is found when they detect a blockage. However, it is essential to have a back-up route identified and ready to be deployed even before the blockage happens. In this case, real-time communication can quickly be resorted when the blockage actually occurs.

\subsection{Our Contributions}
 Given the stringent latency and data rate requirements for mmWave applications, searching for an alternative route \emph{after} blockage can significantly degrade the overall performance. Frequent link blockages for even a few milliseconds affects the end-user's quality of experience (QoE). Thus, it is essential that the blocked link is quickly \emph{detected} and \emph{replaced} with an alternative link. Within this context, designing a multi-hop communication protocol that is robust to blockage and provides an extended range of communication, is of utmost importance to enable pervasiveness mmWave and mesh technologies. Compared with the previous works, this paper focuses on investigating the performance of multi-hop communications and developing a hop-by-hop multi-path routing protocol for mmWave systems.


\section{Hop-by-Hop Multi-Path Routing}
\label{sec:orbit}

We consider a mmWave network that consists of several STAs such that the intermediate STAs are able to relay data traffic from the originating STA to the destination STA (depending on the connectivity and links configurations between STAs). Each directional link is associated with a cost metric that captures the quality of that link. Cost of a route is defined as the summation over the cost of individual constituent links of the route.  The goal is to find one primary and one backup link from the originating STA to neighbor STAs to achieve a \emph{hop-by-hop multi-path routing}. In hop-by-hop multi-path routing, the routing table at each STA contains two next-hops per destination (not for the entire route). If the link to the primary next-hop is blocked, the STA switches to its backup next-hop. In \cite{Schneider2019HopbyHopMR}, hop-by-hop routing  is proposed for the Internet data packets that can be split between multiple next hop switches. This in turn improves the efficiency and robustness. In this paper, hop-by-hop routing is considered for mmWave systems in order to mitigate blockage issues. In traditional multi-path AODV protocols (e.g., \cite{LAI2007453,park2008}), when the primary path from the source to destination is unavailable, the source node switches to the \emph{next alternate end-to-end path}. In contrast, the hop-by-hop multi-path routing adds a local repair capability to each node (source and intermediate) by switching to an alternative link for the blocked link, without the need for an entire route change by the source. This may result in a transient non-optimal end-to-end route, which can be fixed in the next round of routing tables reset. 

\begin{figure}[ht]
    \vspace{-.5cm}
	\centering
	\includegraphics[scale=.6]{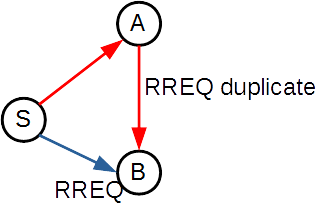}
	\caption{An example of RREQ and its duplicate with three STAs}
	\label{fig:duplicate-rreq}
\end{figure}

\textbf{Route Discovery Messages:} In order to disseminate
the routing control messages across the network, either omni-directional transmissions or directional signals can be used. In the former, due to the short-range nature of mmWave signals, omni-directional signaling may not reach to the neighbor nodes. In the latter, directional transmissions incur high signaling
overhead to disseminate control messages since the same message needs to be transmitted across different sectors to cover all the neighbor nodes. In this paper, we assume that the STAs have performed the sector sweep (SSW) procedure with the 1-hop neighbors. In order to establish a multi-hop route, the originating STA sends directional RREQ frames to its neighbor STAs.  Each 1-hop neighbor  receives the RREQ frame and updates its reverse route  to the originating STA. Each neighbor STA then forwards the RREQ to its 1-hop neighbor as well, excluding the transmitter STA from which the RREQ was received. As the forwarding continues, intermediate STAs may receive duplicate RREQ from other STAs. For example, consider a network with three STAs as shown in Fig. \ref{fig:duplicate-rreq}. In this example, STA B receives two RREQ frames: one from the originating STA S and one through STA A. 
\begin{figure}[t]
     \centering
     \begin{subfigure}[b]{0.14\textwidth}
        \centering
        \includegraphics[scale=.5]{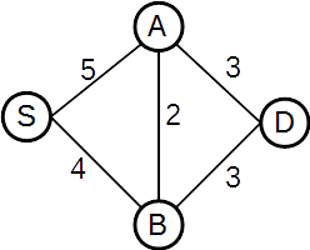}
        \caption{}
        \label{fig:network}
     \end{subfigure}
     ~
     \begin{subfigure}[b]{0.14\textwidth}
        \centering
        \includegraphics[scale=.30]{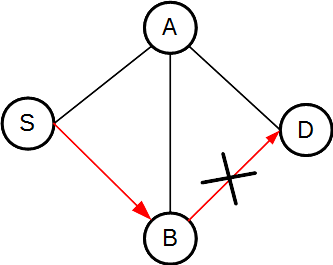}
        \caption{}
		\label{fig:B-D_blocked}
 	\end{subfigure}
    ~
     \begin{subfigure}[b]{0.14\textwidth}
        \centering
        \includegraphics[scale=.29]{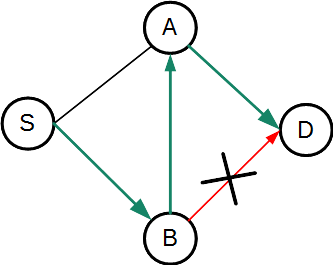}
        \caption{}
		\label{fig:B-D_backup}
     \end{subfigure}
    
\vspace{-.3cm}

    \caption[=]{\small{Network topology and blocked link that causes the backup link from STA B to A to be deployed in order to relay the data traffic from STA S to STA D. }}
        
\end{figure}


By receiving RREQ messages, the best RREQ and second best RREQ frames (in terms of cost) determine the next-hop and backup next-hop node to the originating STA in the routing table of the intermediate STAs. In the example above, STA B sets STA A as the backup next-hop to reach to STA S, assuming that the link metric from S to B is better than sum of the link metrics S to A and A to B.  In order to reduce the routing overhead, the intermediate STA picks the best received RREQ, and forwards it to its neighbor STAs. To avoid looping, the STA records the forwarding action in its Forwarding Table. 

 Destination STA receives potentially several RREQ messages, and sends an RREP frame to the same STA from which an RREQ was received. Each intermediate STA that receives an RREP message, updates its routing table to the destination STA. 
If the intermediate STA receives more than one RREP, it picks the best RREP frame and forwards it to its 1-hop neighbor STAs. Intermediate STA records the forwarding operation in its Forwarding Table. Similar to RREQ, each RREP frame and its duplicate versions determine the next-hop and backup next-hop. The process of forwarding RREP continues until the RREP message is received at the originating STA. Originating STA potentially receives more than one RREP message. It picks the best and second best RREP message and records them as the next-hop and backup next-hop to reach to the destination STA. Each STA proactively makes sure its routing table entries are up-to-date and two next-hop options are reachable. 

 \begin{figure*}[t]
     \centering
     \begin{subfigure}[b]{0.32\textwidth}
        \centering
        \includegraphics[scale=.48,trim=0cm 1.5cm 2.5cm 0cm, clip]{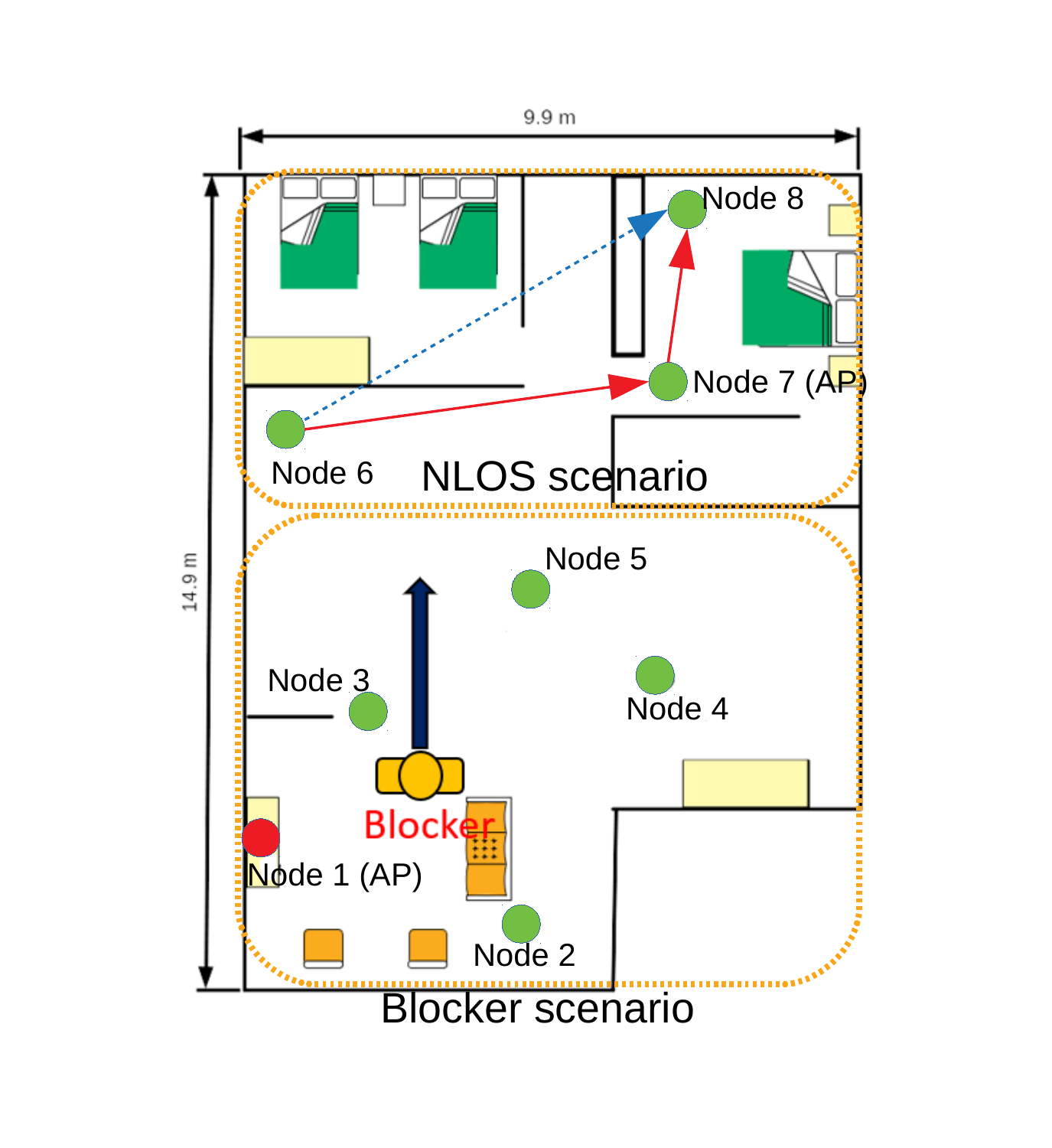}
        \caption{Location of the nodes with human blockage}
        \label{fig:room}
     \end{subfigure}
     \hspace{.25cm}
     \begin{subfigure}[b]{0.32\textwidth}
        \centering
        \includegraphics[scale=.6,trim=2cm 2cm 2cm 0cm, clip]{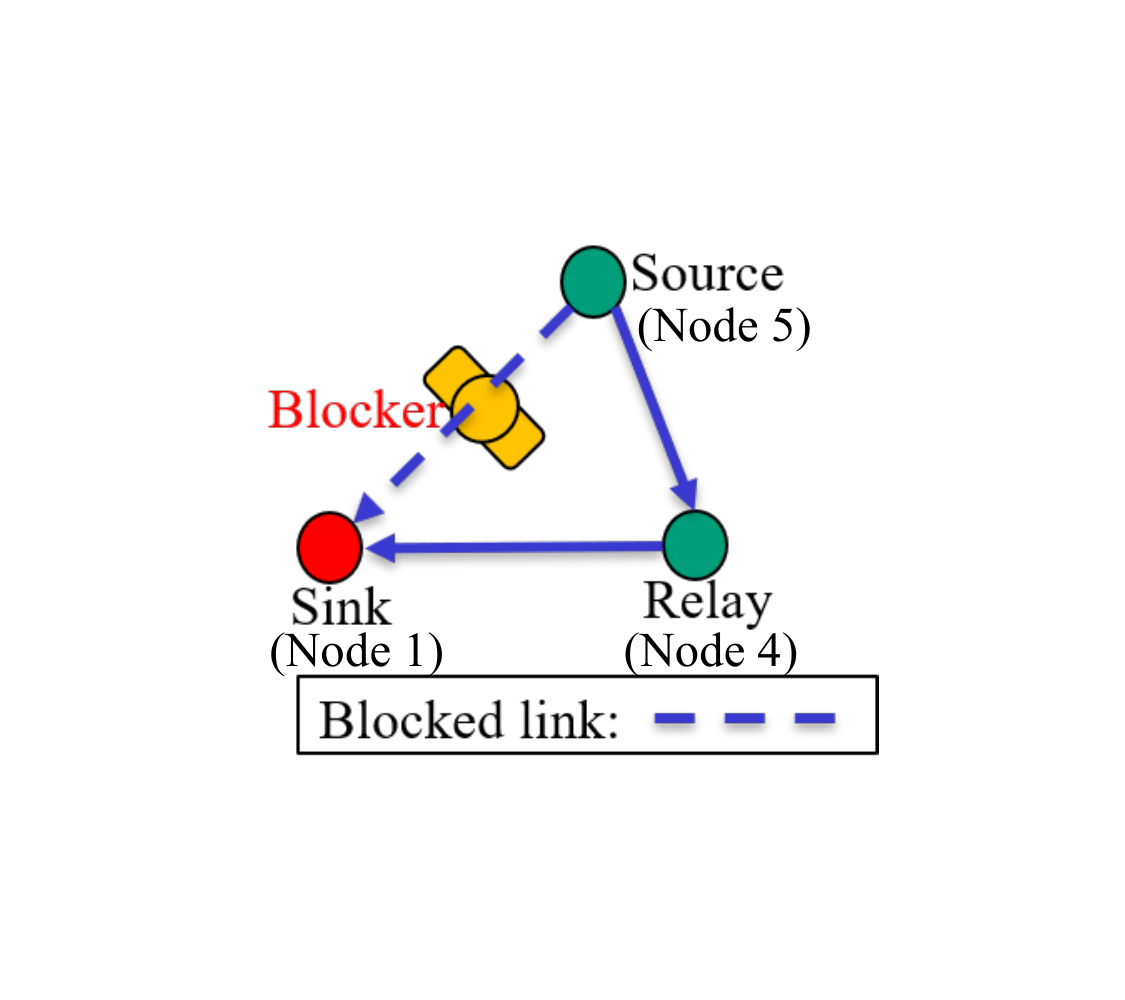}
        \caption{Multi-hop topology}
		\label{fig:topology}
 	\end{subfigure}
     \begin{subfigure}[b]{0.32\textwidth}
        \centering
        \includegraphics[scale=.32,trim=0cm 0cm 0cm 0cm, clip]{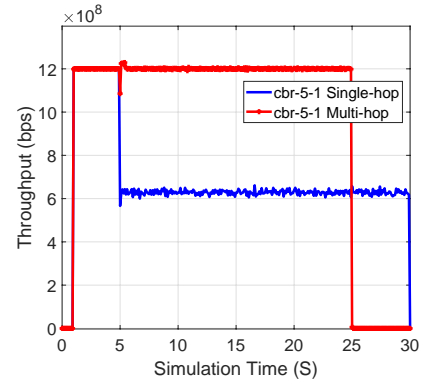}
        \caption{Throughput of multi-hop vs. single-hop}
		\label{fig:C211_throughput}
     \end{subfigure}
    
    \caption[=]{\small{An indoor mmWave network with three nodes. Node 5 is the source and node 1 is the destination for 1.2 Gbps data traffic starting at 1 second and ending at 25 seconds. Node 4 is the relay node.}}
        
\end{figure*}
 
 To demonstrate this process, let us consider the network topology depicted in Fig. \ref{fig:network} where STA S is the source, STA D is the destination, and STAs A and B serve as relay nodes. Weight of each edge determines the cost of that directional link. In this case, the routing table at STA B has the primary next-hop of D to reach STA D (with the cost of 3), while STA A is listed as the backup next-hop for STA B to reach STA D with the cost of 5.  The routing table at STA B is shown in Table \ref{tab:routing-sta-S}.
 Now, assume that the direct link from B to D gets blocked as shown in Fig. \ref{fig:B-D_blocked}. In this case, based on the routing table, STA B switches to its backup next-hop (STA A) that is ready to be deployed. The updated network topology is shown in Fig. \ref{fig:B-D_backup}. After the local repair step at STA B, it transmits an error message to its neighbors so that they can correct their routes to converge to a global optimal route. Temporarily switching to a backup link at STA B is faster than multi-path AODV that would typically require switching to the alternative end-to-end route S$\rightarrow$A$\rightarrow$D \cite{park2008}.  
 \vspace{-.2cm}
  \begin{table}[ht]
 	\centering
 	 	\caption{\small{Routing table at B with primary and backup next-hops.}}
 	\hspace{-.7cm}
 	\begin{tabular}{@{}lccc@{}}\toprule
 		\textbf{Destination} & \textbf{STA S} & \textbf{STA A} & \textbf{STA D }
 		\\ \midrule
 		Next hop & S & A & D  
 		\\
 		Route metric & 4 & 2 & 3  \\
 		Backup next-hop & A & NA & A \\
 		Backup route metric & 7 & NA & 5 \\
 		\bottomrule
 	\end{tabular}
 	\label{tab:routing-sta-S}
 \end{table}
 

\textbf{Route Discovery Triggers:}  We use three mechanisms to detect that a neighbor STA is no longer available.
First, similar to the AODV protocol, we use HELLO messages such that if a neighbor STA misses HELLO transmissions, that neighbor STA is assumed to be not available. Second, if the number of transmissions for a frame exceeds a threshold, the link is assumed to be broken. This threshold is set to 7 for short frames and 4 for long frames. Third, we implement a trigger based on the modulation and coding scheme (MCS) index such that if the MCS index drops below a specified value\footnote{We assume that an adaptive rate controller such as ARF is deployed.}, we consider that link not useful, and route discovery process (i.e., sending RREQ) is initiated. This in turn, can guarantee a minimum throughput at the destination STA.

\section{Simulation Results}
\label{sec:simulation}
In this section, we present simulation results to demonstrate the benefits of multi-hop communications for indoor 60 GHz systems such as AR/VR and 8K UHD wireless streaming.  The main contribution of this paper is to investigate the performance of multi-hop mmWave for indoor scenarios, rather than improving the performance of AODV-type protocols. 
 \subsection{Setup}
We simulate a mmWave network under various scenarios where all nodes are equipped with the IEEE 802.11ad PHY and MAC specifications.  The IEEE 802.11ad SC PHY with AWGN channel model is used such that the simulation parameters are summarized in Table \ref{tab:parameter}. In order to model the propagation environment, we use Remcom X3D ray tracer with High Fidelity Propagation Model (HFPM) enabled. Materials used for simulating the room environment along with their properties are listed in Table \ref{tab:material}.  In the simulations, we set the total number of computed paths to be 25; number of reflections 3, number of diffraction 1, and number of transmissions 3.  In order to assess the performance of multi-hop communications, we measure the end-to-end throughput, delay, and cumulative distribution function (CDF) for delay. Throughput is defined as the total number of bits received divided by the simulation time. The end-to-end delay is measured from the time that a data  packet is generated at the source node until it is received at the destination node. This quantity also captures the latency introduced by the relay node, noting that the main source of delay is due to the channel access delay and queuing delay at the source and relay. 
\begin{table}[t]
	\caption{\small{Simulation parameters}}
 	\centering
 	\vspace{-.2cm}
  	\begin{tabular}{@{}lc@{}}\toprule
  	\textbf{Simulation Parameter} & \textbf{Value} \\ \midrule
  	Transmit power & 18dBm  \\
      Preamble detection threshold & -68dBm \\
      Noise level & -70.6dBm  \\
      Energy detection threshold & -48dBm  \\
      Channel access scheme & Contention based \\
      Beacon interval (BI) & 100ms  \\
      Beacon header interval (BHI) & 5ms  \\
      Data transmission interval (DTI) & 95ms  \\
      Rate controller & ARF \\
      Maximum number of aggregated MPDU & 64 \\
      Transmit opportunity duration (TXOP) & 300 $\mu$s \\
      Human blocker path loss & 20 dB \\
      Human blocker dimensions (length, width, height) & (0.5m, 0.5m, 1.8m) \\
 	\bottomrule
 	\end{tabular}
 	\label{tab:parameter}
 \end{table}
\begin{table}[t]
 	\caption{\small{Materials characteristics  used in the simulations}}
\begin{threeparttable}
 	\centering
 		\vspace{-.2cm}
  	\begin{tabular}{@{}lccc@{}}\toprule
  	\textbf{Material} &  \textbf{Relative Permittivity\tnote{1}} & \textbf{Conductivity\tnote{2}} & \textbf{Thickness~(m)} \\ \midrule
 Brick wall & 5.31 & 0.8967 & 0.3\\
  Concrete wall & 5.31 & 0.8967 & 0.3 \\
  Wood & 1.99 & 0.3784 & 0.03\\
  Glass & 6.27 & 0.5674 & 0.001\\
	\bottomrule
 	\end{tabular}
 \begin{tablenotes}
       \item[1]Relative permittivity with respect to free space or vacuum.
       \item[2]Conductivity is measured in terms of Siemens per meter (S/m). 
     \end{tablenotes}
 \end{threeparttable}
 	\label{tab:material}
 \end{table}

 \subsection{Blocker Scenario with a Single Data Flow}

\begin{figure*}[t]
     \centering
     \vspace{-.5cm}

     \begin{subfigure}[b]{0.23\textwidth}
        \centering
        \includegraphics[scale=.25,trim=0cm 0cm 1.3cm 0cm,clip]{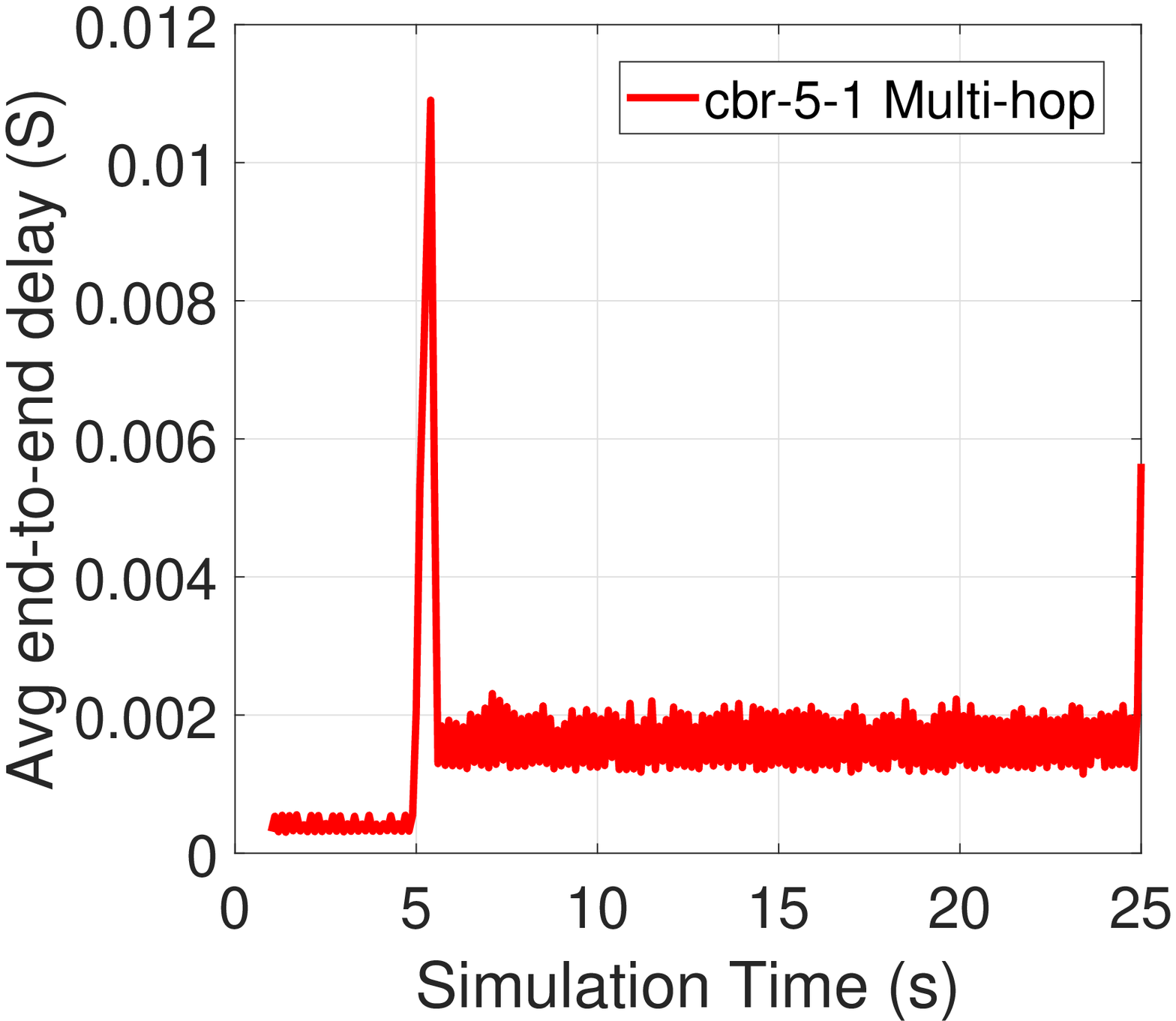}
        \caption{ Multi-hop delay }
        \label{fig:C211-delay}
     \end{subfigure}
     ~
     \begin{subfigure}[b]{0.23\textwidth}
        \centering
        \includegraphics[scale=.25,trim=0cm 0cm 1.3cm 0cm, clip]{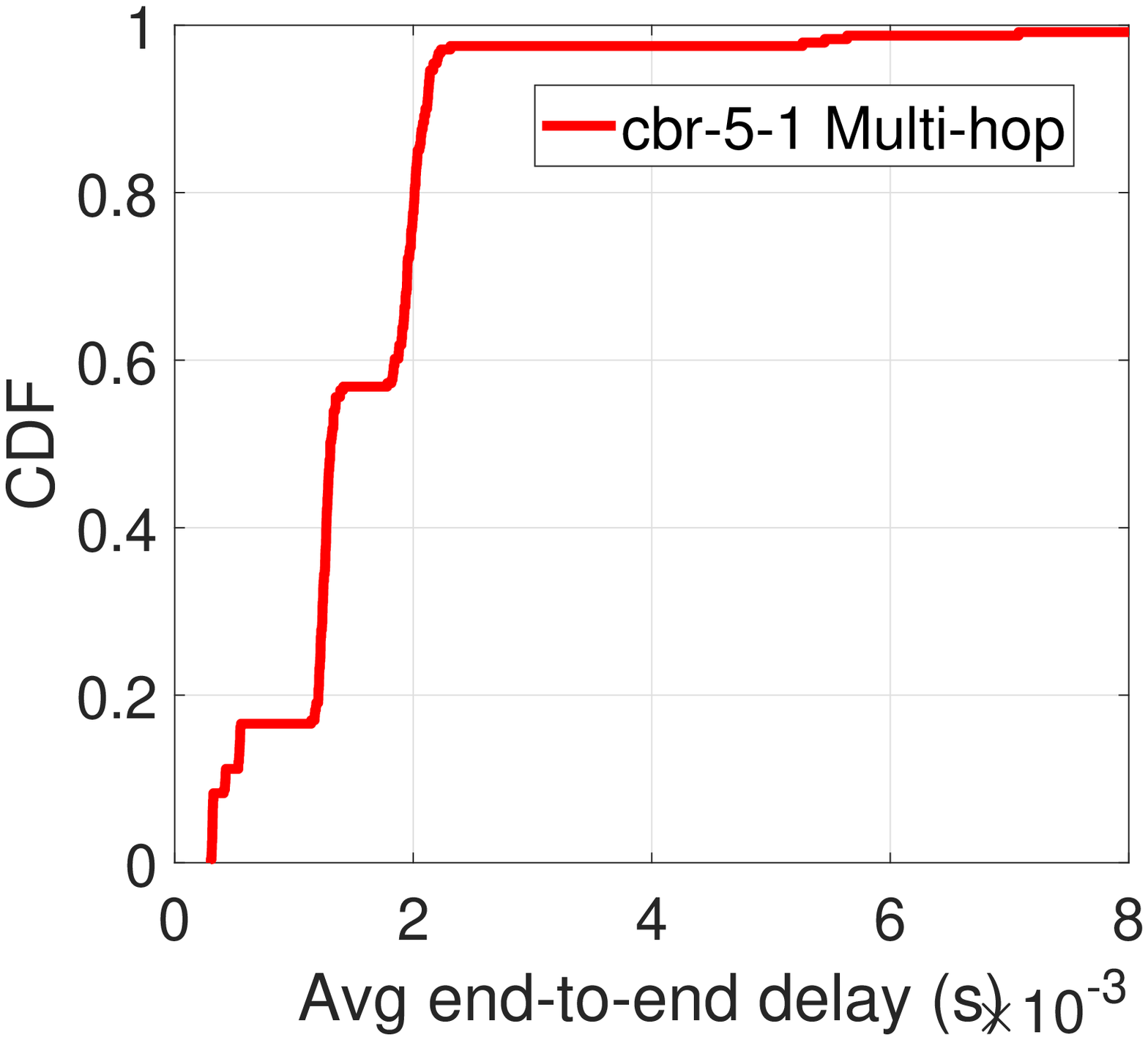}
        \caption{ CDF for multi-hop delay }
        \label{fig:C211-delay-CDF}
 	\end{subfigure}
    ~
     \begin{subfigure}[b]{0.23\textwidth}
        \centering
        \includegraphics[scale=.25,trim=0cm 0cm 1.3cm 0cm, clip]{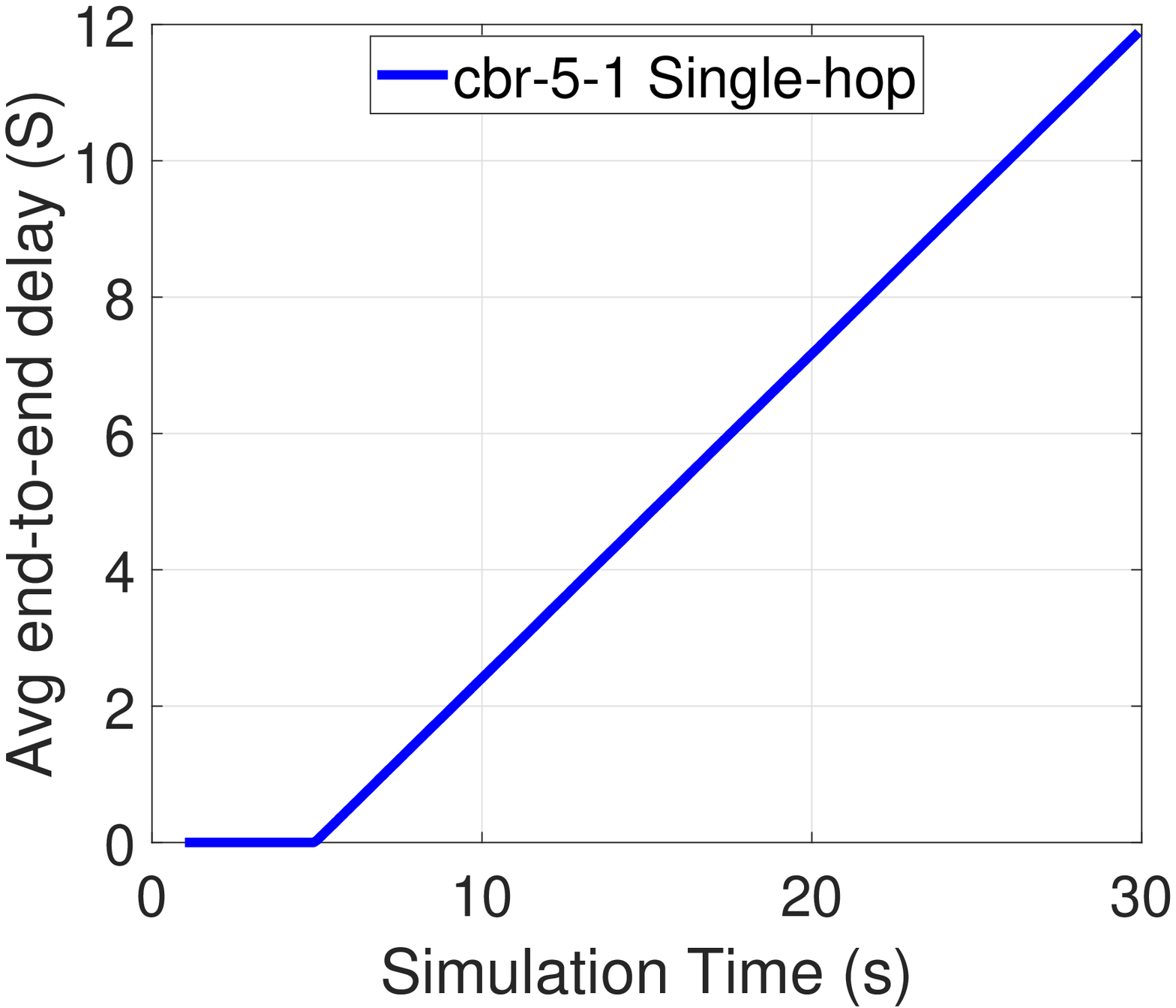}
        \caption{ Single-hop delay }
 		\label{fig:C211-delay-single-hop}
     \end{subfigure}
    ~
     \begin{subfigure}[b]{0.23\textwidth}
         \centering
         \includegraphics[scale=.25,trim=0cm 0cm 1cm 0cm, clip]{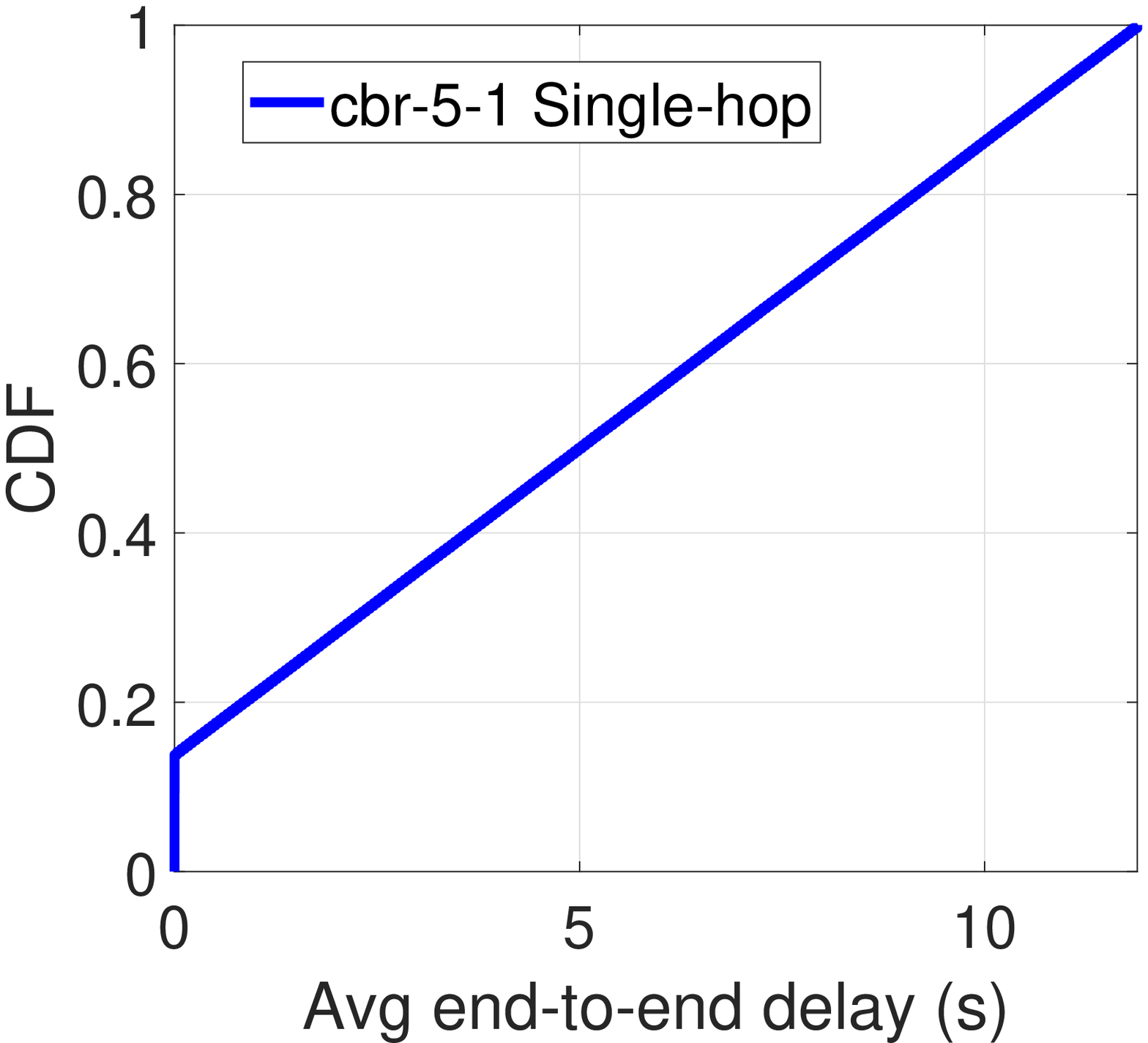}

         \caption{ CDF for single-hop delay }
         \label{fig:C211-delay-single-hop-CDF}
     \end{subfigure}
\vspace{-.2cm}

    \caption[=]{\small{Multi-hop vs single-hop delay performance under blocker scenario with one constant bit rate (CBR) data flow from 5 to 1 (cbr-5-1). }}
        
\end{figure*}
 We start with a mmWave network with three nodes 1, 4, and 5 (as shown in Fig. \ref{fig:room}) where node 5 is the source and node 1 (AP) is the destination of 1.2 Gbps constant bit rate (CBR) traffic. For this set of simulation, other nodes are inactive. 
 At time 5s, the link between the source and destination (sink) node is blocked by a human body that is modeled as an additional $20$ dB path loss. Human body dimensions are set to  0.5m in length, 0.5m in width, and 1.8m in height.  Figure \ref{fig:topology} graphically shows the network topology and blocking object. To reach to node 1, the routing table at node 5 has node 1 as the primary next-hop and node 4 as the backup next-hop.  Figure \ref{fig:C211_throughput} shows the throughput performance of single-hop and multi-hop communications under blockage. From the results, we notice that when blockage starts single-hop link uses a lower MCS and thus throughput drops. On the other hand,  multi-hop topology is able to maintain the high-throughput performance by exploiting the relay node. Figure \ref{fig:C211-delay} through \ref{fig:C211-delay-single-hop-CDF} show the end-to-end delay and CDF for the delay under single-hop and multi-hop topologies. Multi-hop delay performance is considerably better than single-hop delay, with the $99^{\text{th}}$ percentile of less than 5ms. Simulation results show that after blockage at 5s, the multi-hop protocol adds an alternative route at $5.012$s.  

 \vspace{-.1cm}
\subsection{Blocker Scenario with Multiple Data Flows}
\begin{figure}[t]
 	\centering
 	\vspace{-.5cm}
 		\includegraphics[scale=.25,trim=0cm 0cm 0cm 0cm, clip]{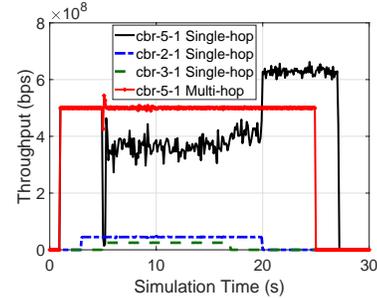}
 		\vspace{-.2cm}
 	\caption{\small{Throughput comparison of multi-hop vs. single-hop}}
 	\label{fig:C511_throughput}
\end{figure}

\begin{figure}[t]
  \begin{subfigure}[t]{.2\textwidth}
    \centering
    \includegraphics[scale=.22,trim=0cm 0cm 0cm 0cm, clip]{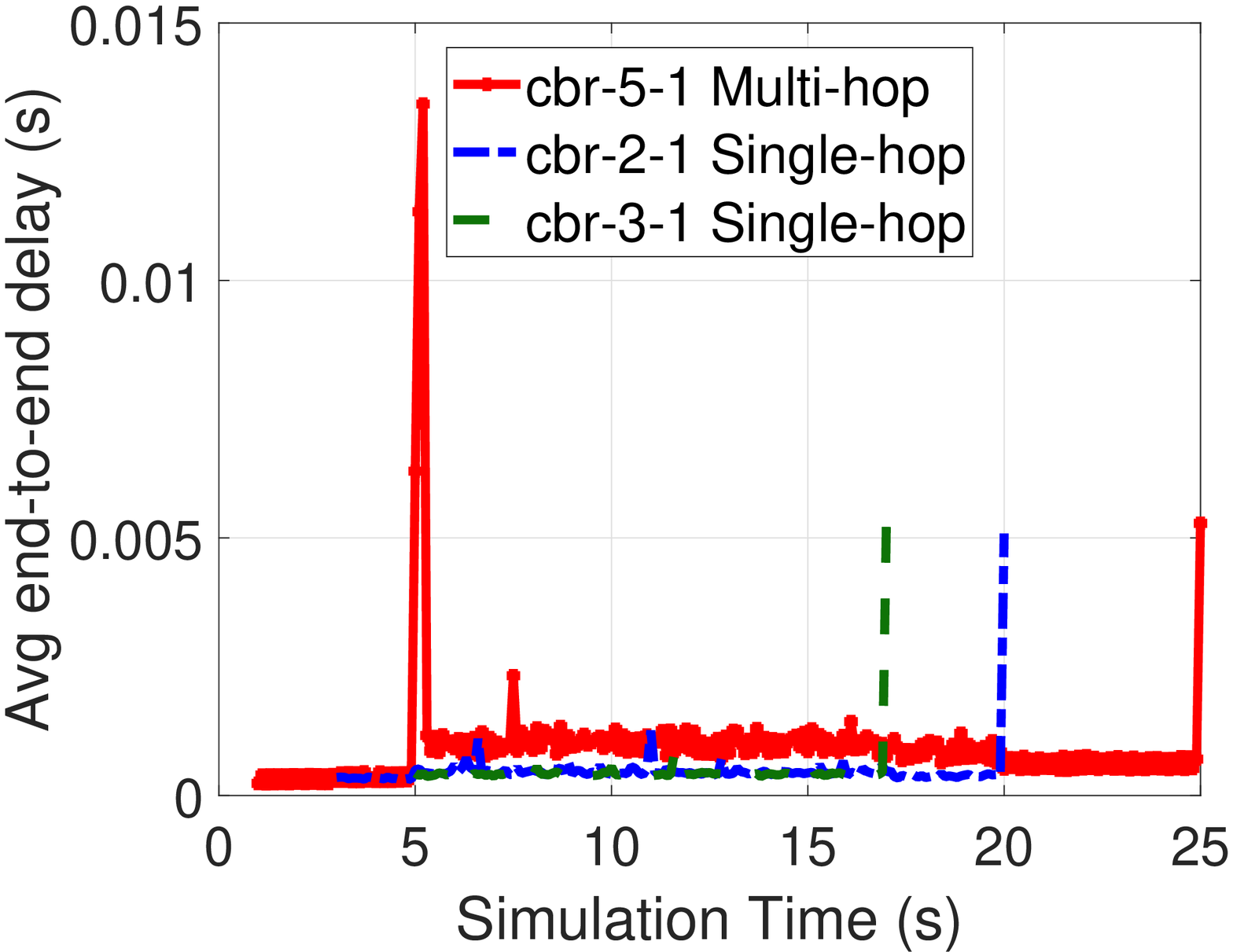}
    \caption{Multi-hop delay}
    \label{fig:C511-delay}
  \end{subfigure}
  \begin{subfigure}[t]{.2\textwidth}
  \hspace{-1cm}
    \centering
   \includegraphics[scale=.22,trim=0cm 0cm 0cm 0cm, clip]{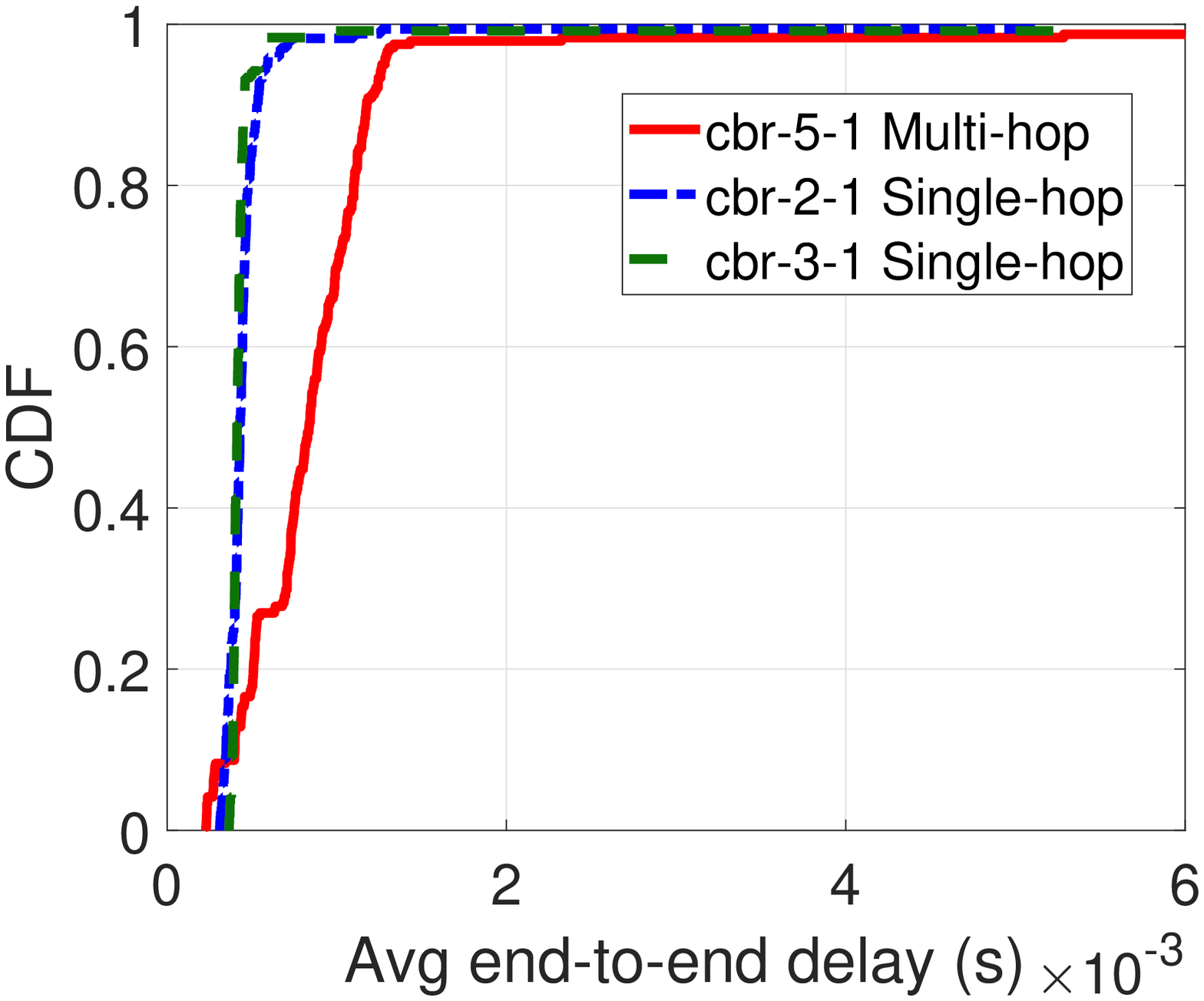}
    \caption{CDF for multi-hop delay}
    \label{fig:C511-delay-CDF}
  \end{subfigure}
  \begin{subfigure}[t]{.2\textwidth}
    \centering
    \includegraphics[scale=.22,trim=0cm 0cm 0cm 0cm, clip]{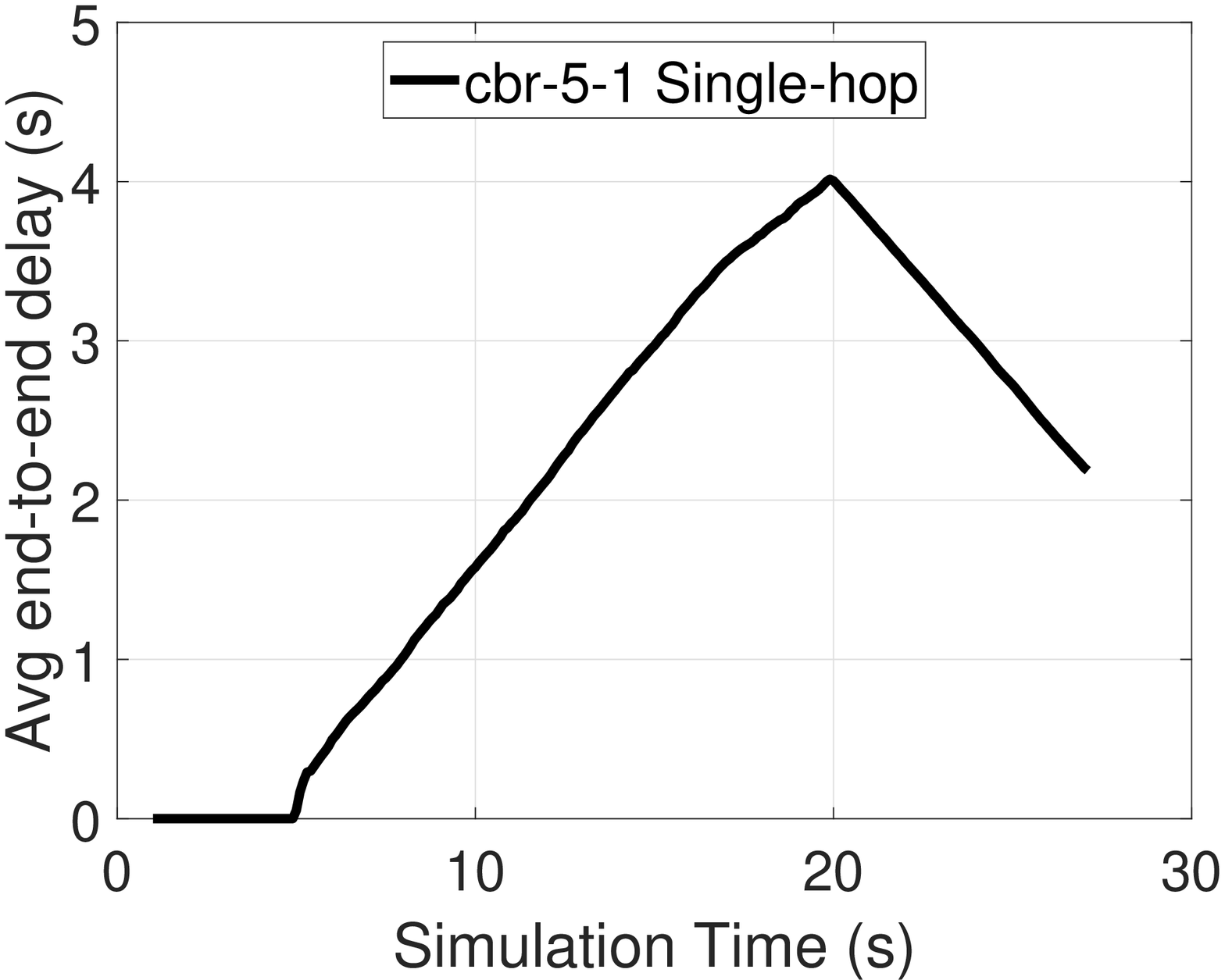}
    \caption{Single-hop delay}
    \label{fig:C511-delay-single-hop}
  \end{subfigure}
  \hspace{.4cm}
  \begin{subfigure}[t]{.2\textwidth}
    \centering
    \includegraphics[scale=.22,trim=0cm 0cm 0cm 0cm, clip]{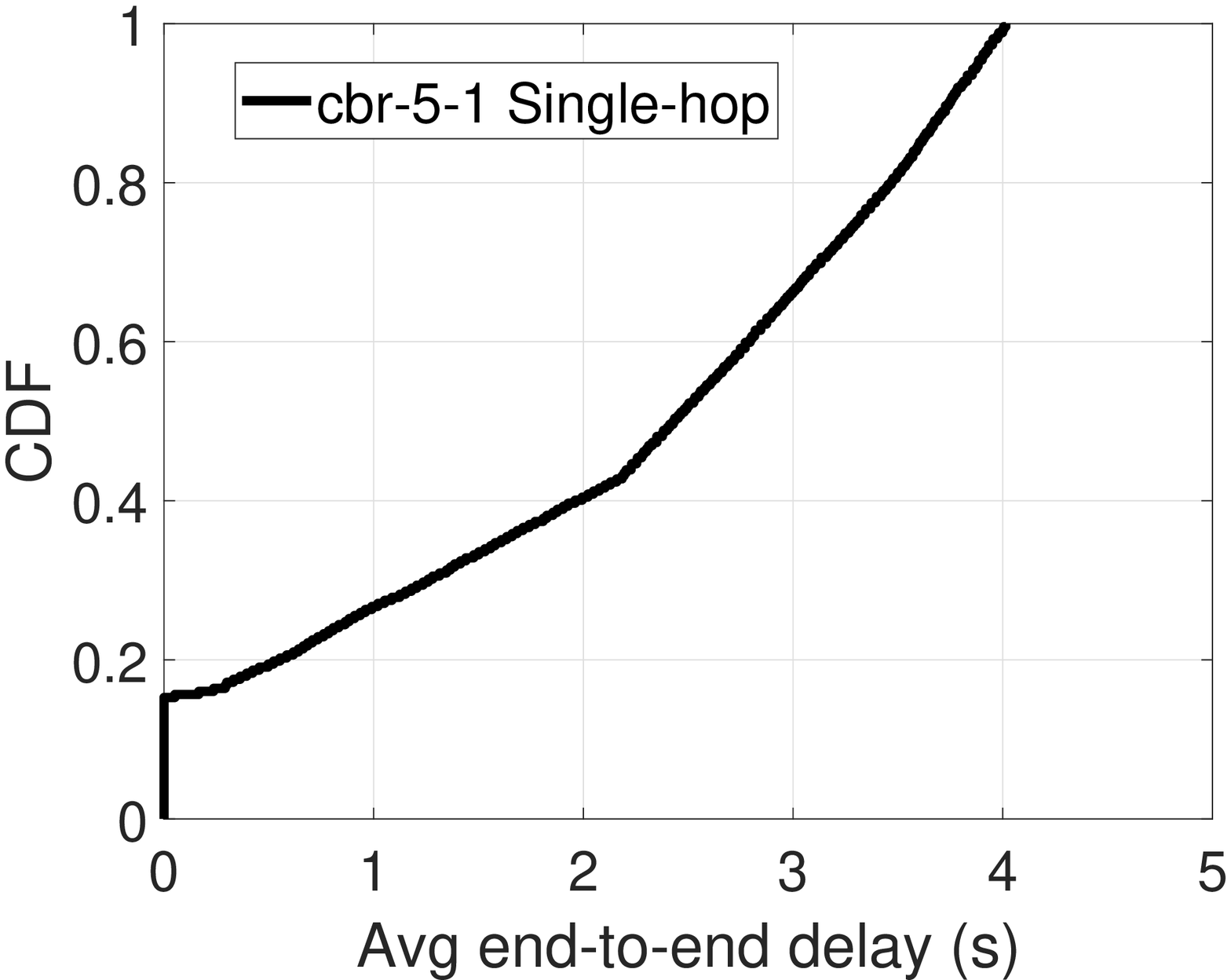}
    \caption{CDF for single-hop delay}
    \label{fig:C511-delay-single-hop-CDF}
  \end{subfigure}
  \caption{\small{Delay performance of multi-hop vs single-hop under blocker scenario with multiple CBR data flows.}}
  \label{fig:C511_delay}
\end{figure}

Next, we activate data flows from node 3 and node 2 to node 1 (AP). The CBR data rates are reduced to be within the capacity region, and are set to 500 Mbps from 5 to 1, 45 Mbps from 2 to 1, and 25 Mbps from 3 to 1. Figure \ref{fig:C511_throughput} shows the throughput performance of single-hop and multi-hop networks. In the presence of other data flows, single-hop topology does not provide a reliable and stable throughput even at a lower data rate. On the other hand, multi-hop topology enhances the throughput performance.  In addition, Fig. \ref{fig:C511-delay} through \ref{fig:C511-delay-single-hop-CDF} compare the delay performance of multi-hop vs single-hop for different data flows. From the results, we observe that multi-hop delay for CBR 5 to 1 is much smaller than single-hop delay, and that the average multi-hop latency is less 5ms.

\subsection{Non-Line-of-Sight (NLOS) Scenario}
\begin{figure}[t]
	\begin{subfigure}[t]{.5\textwidth}
		\centering
		\includegraphics[scale=.25,trim=0cm 0cm 0cm 0cm, clip]{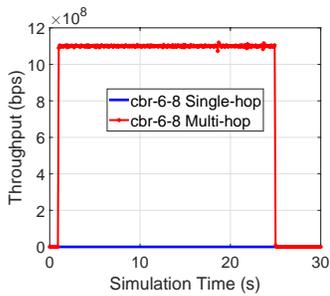}
		\vspace{-.2cm}
		\caption{Throughput performance of multi-hop vs single-hop}
		\label{fig:C411-Tput}
	\end{subfigure}
	\begin{subfigure}[t]{.5\textwidth}
		\centering
		\includegraphics[scale=.238,trim=1cm 0cm 0cm 0cm, clip]{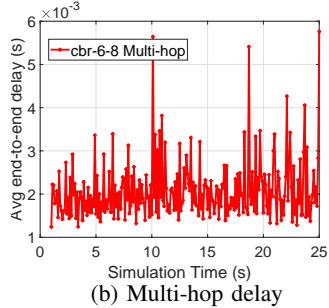}
		\vspace{-.2cm}
		\caption{Multi-hop delay}
		\label{fig:C411-delay}
	\end{subfigure}
	\begin{subfigure}[t]{.5\textwidth}
		\centering
		\includegraphics[scale=.235,trim=0cm 0cm 0cm 0cm, clip]{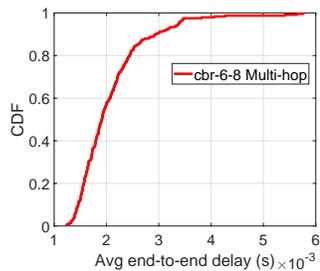}
		\vspace{-.2cm}
		\caption{CDF for multi-hop delay}
		\label{fig:C411-delay-CDF}
	\end{subfigure}
	\vspace{-.3cm}
	\caption{\small{Performance of multi-hop vs single-hop for NLOS}}
	\label{fig:C511_delay2}
	\vspace{-0.5cm}
\end{figure}

In this set of simulations, we examine the performance of multi-hop routing for range extension and NLOS scenarios. As shown in the \emph{NLOS scenario} in Fig. \ref{fig:room}, we consider a scenario such that there is no direct path between the source node 6 and destination node 8. Data traffic generated at the source is 1.1 Gbps.
From the simulation results shown in Fig. \ref{fig:C411-Tput}, it is clear that direct communication between the source and destination is not able to provide sufficient link budget, and thus throughput is zero. On the other hand, multi-hop communication using the relay node is able to provide high throughput.
 Delay performance of multi-hop is shown in Fig. \ref{fig:C411-delay} and Fig. \ref{fig:C411-delay-CDF} from which we observe that the latency is almost always less than 5ms.  
 \vspace{-.2cm}
\section{Conclusion}
\label{sec:conclusion}
In this paper, we investigated the benefits of network layer solutions and on-demand routing protocol with backup routes  
to ensure reliable and robust mmWave communication under severe conditions (blockage and NLOS). 
Our hop-by-hop multi-path routing protocol establishes one primary and one reserved link per destination such that once the primary link is blocked, the backup link is ready to be deployed.  To verify the performance of our protocol, we conducted  system-level simulations based on the IEEE 802.11ad PHY and MAC specifications. Our simulations confirm the validity of our approach to sustain high throughput and low latency performance under blockage and NLOS scenarios. 

\vspace{-.2cm}
\small{
\bibliographystyle{ieeetr}
\bibliography{Ref-Hashemi}}

\end{document}